\documentstyle[12pt]{article}
\input epsfig.sty
\def\beq{\begin{equation}}
\def\eeq{\end{equation}}
\def\bea{\begin{eqnarray}}
\def\eea{\end{eqnarray}}

\def\NP{{\it Nucl. Phys.} }

\def\PRL{{\it Phys. Rev. Lett.} }
\def\ap{\alpha ^{\prime}}
\begin {document}
\begin{titlepage}
Revised, September 1998 \\
\begin{flushright}
HU Berlin-EP-98/40\\
\end{flushright}
\mbox{ }  \hfill hepth@xxx/9807093
\vspace{5ex}
\Large
\begin {center}
\bf{$Q\bar Q$ potential from AdS-CFT relation at $T\geq 0$:\\
Dependence on orientation in internal space and higher curvature corrections}
\end {center}
\large
\vspace{1ex}
\begin{center}
H. Dorn, H.-J. Otto
\footnote{e-mail: dorn@physik.hu-berlin.de otto@physik.hu-berlin.de}
\end{center}
\normalsize
\it
\vspace{1ex}
\begin{center}
Humboldt--Universit\"at zu Berlin \\
Institut f\"ur Physik, Theorie der Elementarteilchen \\
Invalidenstra\ss e 110, D-10115 Berlin
\end{center}
\vspace{4ex}
\rm
\begin{center}
{\bf Abstract}
\end{center}
Within the classical approximation we calculate the static $Q\bar Q$ potential via the AdS/CFT relation for nonzero temperature and arbitrary internal orientation of the quarks. We use a higher order curvature corrected target space background. For timelike Wilson loops there arises a critical line in the
orien\-tation-dis\-tance plane which is shifted to larger distances relative to the calculation with uncorrected background. Beyond that line there is no $Q\bar Q$-force. The overall vanishing of the force for antipodal orientation known from zero temperature remains valid.
The spacelike Wilson loops yield a string tension for a (2+1)-dimensional
gauge theory, independent of the relative internal orientation, but sensitive
to the background correction.
\vfill      
\end{titlepage}
%%%%%%%%%%%%%%%%%%%%
\setcounter{page}{1}
\pagestyle{plain}
\section{Introduction}
Part of the recent interest in relations between gauge field theories and
string theory/super\-gravity on certain nontrivial backgrounds \cite{malda0,witten0,kleb} is due to a conjecture expressing the Wilson loops in the gauge theory by the partition function of a string fulfilling boundary conditions set
by the geometry of the Wilson loop \cite{malda,rey,witten}. This has been used
to calculate the heavy quark-antiquark potential for ${\cal N}=4$ super
Yang-Mills in 3+1 dimensions in the t'Hooft limit and large effective coupling
from the classical string action in $\mbox{AdS}_5\times S^5$ \cite{malda,rey}.
The procedure has been extended to general dimensions and to geometries arising as some limit of near extremal D-brane configurations \cite{malda,witten,theisen,brand}. These background configurations with compactified Euclidean time
coordinate are related to equilibrium $T>0$ field theory. Since the periodicity condition breaks supersymmetry one gains access to the quark-antiquark
potential in non-supersymmetric gauge theories (hopefully QCD) in one dimension lower \cite{witten,theisen,brand}. 

The present paper will be concerned mainly with two aspects of this kind of calculations. In the $\mbox{AdS}_5\times S^5$ calculation of ref.\cite{malda} a Coulombic potential arises, which is switched off for quark-antiquarks 
oriented antipodally with respect to the internal $S^5$. This is in agreement with an argument telling that the corresponding configuration is of BPS type
\cite{callan,malda}. We would like to know what happens in this antipodal
case for $T>0$, where supersymmetry is broken.

The classical approximation used on the string side of the duality is justified in the t'Hooft limit for large effective coupling, corresponding to small curvature of $\mbox{AdS}_5\times S^5$ or of the relevant region
of the near extremal D-brane configuration for $T>0$. A complete treatment
has to include both quantum corrections and corrections of the target space
background, which after all has to be a solution for the effective string action exact in $\ap $. As a first step in this direction we perform the classical calculation using the next order in $\ap $ corrected background of ref.\cite{tsey,paw}. At $T>0$ the quark-antiquark potential vanishes beyond some critical distance \cite{theisen,brand}. This total screening is expected to be weakened if
the corrections just mentioned are taken into account \cite{gross}. Therefore, our focus will be on the behaviour of the critical line in the $(\Delta\Theta ,L)$-plane (relative $S^5$ orientation and distance), if the correction of \cite{tsey,paw} is switched on.\\     

Our 10D metric for non-zero temperature $T>0$ is given by
\bea
ds^2&=&\ap\left \{\frac{U^2}{R^2}(1-\frac{U_T^4}{U^4})
~e^{\gamma A(\frac{U_T}{U})}~(dx^0)^2+\frac{U^2}{R^2}~e^{\gamma C(\frac{U_T}{U})}~(dx^i)^2\right .\nonumber\\
&&\left .+\frac{R^2}{U^2(1-\frac{U_T^4}{U^4})}~e^{\gamma B(\frac{U_T}{U})}~(dU)^2
+R^2~e^{\gamma D(\frac{U_T}{U})}~(d\Omega _5)^2\right \}~=~G_{MN}dx^Mdx^N~,
\label{1}
\eea
with
\footnote{The dimensionless parameter $\frac{1}{R}$ controls the curvature
and is related to Yang-Mills quantities via $R^2=\sqrt{2g_{YM}^2N}$, which is equal to the effective coupling in the t'Hooft limit.}
\beq
U_T~=~\pi R^2~T~(1+15\gamma )^{-1}~,
\label{2}
\eeq
\bea
A(\rho )&=&-75\rho ^4-\frac{1225}{16}\rho ^8+\frac{695}{16}\rho ^{12}~,\nonumber\\
B(\rho )&=&~75\rho ^4+\frac{1175}{16}\rho ^8-\frac{4585}{16}\rho ^{12}~,\nonumber\\
C(\rho )&=&-\frac{25}{16}\rho ^8-\frac{25}{16}\rho ^{12}~,\nonumber\\
D(\rho )&=&~\frac{15}{16}\rho ^8+\frac{15}{16}\rho ^{12}~.
\label{3}
\eea
We consider
\beq
\gamma ~=~0~~~~~\mbox{or}~~~~~\gamma ~=~\frac{1}{8}\zeta (3)~R^{-6}
\label{4}
\eeq
to describe either the near extremal D-3 brane metric used in refs. \cite{theisen, brand} or the corresponding first order higher curvature corrected metric studied in ref. \cite{tsey,paw}. In these target space metrics we calculate the Nambu-Goto action
\beq
S=\frac{1}{2\pi\ap}\int d\tau d\sigma \sqrt {\mbox{det}(G_{MN}\partial _{\mu}
x^M\partial _{\nu}x^N)}
\label{5}
\eeq
for stationary ($\delta S=0$) string world surfaces fulfilling certain boundary conditions at $U\rightarrow\infty$. In a first case, called timelike, the
boundary of the string world sheet at $U\rightarrow\infty$ consists out
of two lines running at
\footnote{$\vec{\Theta}$ with $\vert \vec{\Theta}\vert = 1$ parametrises $S^5$.}
\beq
x^1=\pm \frac{L}{2},~~~x^2=x^3=0,~~~\vec{\Theta}=\vec{\Theta}_{Q}~~\mbox{or}~~\vec{\Theta}_{\bar Q}
\label{6}
\eeq 
parallel to the $x^0$-axis one times around the compactified $x^0$, with
compactification length $1/T$.
In the second case, called spacelike, the boundary lines at $U\rightarrow\infty$ run parallel to the $x^2$-axis at
\beq
x^1=\pm \frac{L}{2},~~~x^0=x^3=0,~~~\vec{\Theta}=\vec{\Theta}_{Q}~~\mbox{or}~~\vec{\Theta}_{\bar Q}~.
\label{7}
\eeq
Via the CFT/AdS conjecture specified for Wilson loops in \cite{malda,rey,witten} one has in classical approximation for case 1 
\beq
S_{\vert \mbox{\scriptsize case 1}}~=~\frac{1}{T}\cdot V_1(L,\Delta \Theta , T)
\label{8}
\eeq
and for case 2 ($Y$ measures the large $x^2$ interval.)
\beq
S_{\vert \mbox{\scriptsize case 2}}~=~Y\cdot V_2(L,\Delta\Theta )~,
\label{9}
\eeq
where $V_1(L,\Delta \Theta , T)$ is the static quark-antiquark potential
in 3-dimensional space at temperature $T$ for separation $L$ and an angle
$\Delta\Theta$ between the internal orientation $\vec{\Theta}_{Q}$
and $\vec{\Theta}_{\bar Q}$ in $S_5$. $V_2(L,\Delta\Theta )$ for $Y,L\gg\frac{1}{T}$ is the potential in 3-dimensional space-time (i.e. 2-dim. space) at zero
temperature \cite{witten}.
%%%%%%%%%%%%%%%%%%%%%%%%%%%%%%%%%%%%%%%%%%%%%
\section{Timelike Wilson lines}
In this section it is convenient to choose the gauge
\beq
\tau ~=~x^0,~~~~~~\sigma ~=~x^1~.
\label{10}
\eeq
Furthermore, due to the symmetry of the problem, it is sufficient to look for
world surfaces with $\partial _{\tau}x^M=\delta _0^M$, $\partial _{\sigma}x^2=
\partial _{\sigma}x^3=0$, only. Then the action becomes ($^{\prime}$ denotes
derivatives with respect to $\sigma $, $\Theta $ is an angle variable on the great circle on $S^5$ passing $\vec{\Theta}_{Q}$ and $\vec{\Theta}_{\bar Q}$.)
\beq
S=\frac{1}{2\pi T}\int _{-\frac{L}{2}}^{+\frac{L}{2}}d\sigma ~e^{\frac{\gamma}{2}A(\frac{U_T}{U})}~\sqrt{\frac{U^4-U_T^4}{R^4}~e^{\gamma C(\frac{U_T}{U})}+U^{\prime \,2}~e^{\gamma B(\frac{U_T}{U})}+\Theta ^{\prime \,2}~\frac{U^4-U_T^4}{U^2}~e^{\gamma D(\frac{U_T}{U})}~}~.
\label{11}
\eeq
The wanted solutions $U(\sigma ),~\Theta (\sigma )$ are most easily found by making use of the conservation laws derived from the Lagrangian ${\cal L}$ ($S=
\frac{1}{2\pi T}\int d\sigma {\cal L}$)
\bea
C_1&=&\frac{\partial {\cal L}}{\partial U^{\prime }}~U^{\prime }+\frac{\partial {\cal L}}{\partial \Theta ^{\prime }}~\Theta ^{\prime }-{\cal L}~,\nonumber\\
C_2&=&\frac{\partial {\cal L}}{\partial \Theta ^{\prime }}~.
\label{12}
\eea
To be as close as possible to ref. \cite{malda} we parametrise the conserved
quantities $C_1$ and $C_2$ by $U_0=U(0)$ and a parameter $l$ via
\footnote{Making use of $U^{\prime}(0)=0$.} 
\bea
C_2&=&U_0~l~,\nonumber\\
C_1&=&-\frac{U_0^2}{R^2}~e^{\frac{\gamma}{2}\{C(\delta )-D(\delta )\}}~\sqrt{(1-\delta ^4)e^{\gamma \{A(\delta )+D(\delta )\}}-l^2}~,
\label{13}
\eea
with
\beq
\delta ~=~\frac{U_T}{U_0}~.
\label{14}
\eeq
Then using the notation
\beq
f_{\gamma}(y,\delta )~=~(y^4-\delta ^4 )~e^{\gamma    \{A(\frac{\delta}{y})+
C(\frac{\delta}{y})-C(\delta )+D(\delta )\}}~-~l^2y^2~e^{\gamma \{C(\frac{\delta}{y})-D(\frac{\delta}{y})+D(\delta )-C(\delta )\}}
\label{14a}
\eeq
straigthforward integration yields
\beq
\sigma (U)~=~\frac{R^2}{U_T}~\delta ~\sqrt{f_{\gamma}(1,\delta )}
~\int _1^{\frac{U}{U_0}}\frac{e^{\frac{\gamma}{2}\{B(\frac{\delta}{y})-
C(\frac{\delta}{y})\}}~dy}{\sqrt{y^4-\delta ^4}~\sqrt{f_{\gamma}(y,\delta )-
f_{\gamma }(1,\delta )}}~,\label{15}
\eeq
\beq
\Theta (U)~=~l~e^{\frac{\gamma}{2}\{D(\delta )-C(\delta )\}}~\int _1^{\frac{U}{U_0}}\frac{y^2~e^{\frac{\gamma}{2}\{B(\frac{\delta}{y})+C(\frac{\delta}{y})-2D(\frac{\delta}{y})\} }~dy}{\sqrt{y^4-\delta ^4}~\sqrt{f_{\gamma}(y,\delta )-
f_{\gamma }(1,\delta )}}~,\label{16}
\eeq
and in particular, using (\ref{2}), for $L=2\sigma (U\rightarrow\infty ),~~\Delta\Theta =2\Theta (U\rightarrow\infty )$
\beq
L\cdot T~=~\frac{2}{\pi}~(1+15\gamma )\delta ~\sqrt{f_{\gamma}(1,\delta)}~ \int _1^{\infty}\frac{e^{\frac{\gamma}{2}\{B(\frac{\delta}{y})-
C(\frac{\delta}{y})\}}~dy}{\sqrt{y^4-\delta ^4}~\sqrt{f_{\gamma}(y,\delta )-
f_{\gamma }(1,\delta )}}~,
\label{17}
\eeq
\beq
\Delta \Theta ~=~2l~e^{\frac{\gamma}{2}\{D(\delta )-C(\delta )\}}~\int _1^{\infty}\frac{y^2~e^{\frac{\gamma}{2}\{B(\frac{\delta}{y})+C(\frac{\delta}{y})-2D(\frac{\delta}{y})\} }~dy}{\sqrt{y^4-\delta ^4}~\sqrt{f_{\gamma}(y,\delta )-f_{\gamma }(1,\delta )}}~.
\label{17a}
\eeq
Inserting the solution into the action functional (\ref{11}) and making use of the conservation laws (\ref{12}),(\ref{13}) again, we get via (\ref{8}) the
static $Q\bar Q$ potential. Due to the behaviour of the integrand at large $U$
it is divergent. With a preliminary cutoff $\Lambda $ it looks like
\beq
V_1^{(\Lambda )}~=~\frac{U_T}{\pi}~\delta ^{-1}~\int _1^{\frac{\Lambda}{U_0}}
\frac{\sqrt{y^4-\delta ^4}~e^{\frac{\gamma}{2}\{2 A(\frac{\delta}{y})+B(\frac{\delta}{y})+C(\frac{\delta}{y})+D(\delta )-C(\delta )\}}~dy}{\sqrt{f_{\gamma}(y,\delta )-f_{\gamma }(1,\delta )} }~.
\label{18}
\eeq

Defining a renormalised potential along the lines of \cite{malda,theisen,brand} requires a comment. The potential $V_1^{(\Lambda )}$ is linearly divergent. Subtracting the divergence yields a renormalised $V_1$, fixed up to a finite additive constant which per se is completely irrelevant. From the experience with
the $\Delta\Theta =0,~\gamma =0$ case \cite{theisen,brand}, we expect that beyond some critical distance the configuration of two strings stretching straightly
from $U=\Lambda$ to the horizon at $U=U_T$ at fixed values $x_1=\pm \frac{L}{2},~\Theta =\pm\frac{\Delta\Theta}{2}$ is energetically favoured in comparison
to our smooth U-shaped configuration (\ref{15}),(\ref{16}). Choosing $2\cdot\frac{\Lambda -U_T}{2\pi}$, i.e. twice the energy stored in a string stretching between $U_T$ and $\Lambda$ as the subtraction term, the competing piecewise 
straight string configuration has been normalised to zero in \cite{theisen,brand}. As a consequence, in the $L$-region with $V_1\geq 0$ the $Q\bar Q$
force vanishes.

To extend this framework to our modified background (\ref{1}), we first of all
have to check, whether the competing piecewise straight string configuration 
(pssc)
\footnote{At $x^1=-\frac{L}{2},~\Theta =-\frac{\Delta\Theta}{2}$ from $U=\Lambda$ to $U=U_T$, then along the horizon to $x^1=+\frac{L}{2},~\Theta =+\frac{\Delta\Theta}{2}$, and then at fixed $x^1,~\Theta$ back to $U=\Lambda $.}
still has an action (\ref{5}) independent of $L$. Since also with (\ref{1}) we have $G_{00}=0$ at $U=U_T$, the determinant of the induced metric
vanishes for the part of the pssc extending along the horizon. Only the pieces
of the pssc running off the horizon contribute:
\beq
V_{\mbox{\scriptsize pssc}}^{(\Lambda )}~=~\frac{U_T}{\pi}~\int _1^{\frac{\Lambda}{U_T}}~e^{\frac{\gamma}{2}\{A(y^{-1})+B(y^{-1})\}}~dy~.
\label{19}
\eeq
For $\gamma =0$ this reduces to $\frac{\Lambda -U_T}{\pi}$.\\
Now we choose $V_{\mbox{\scriptsize pssc}}^{(\Lambda )}$ as the subtraction constant to get a renormalised $V_1$. Then the potential of the competing pssc
is zero again, and our U-shaped configuration has a potential
$$V_1=\lim _{\Lambda\rightarrow\infty}(V_1^{(\Lambda )}-V_{\mbox{\scriptsize pssc}}^{(\Lambda )})~,$$
leading with (\ref{2}) to
\bea
\frac{V_1}{R^2T}&=&(1+15\gamma )^{-1}\bigg \{\delta ^{-1}~\int _1^{\infty}
\left ( 
\frac{\sqrt{y^4-\delta ^4}~e^{\frac{\gamma}{2}\{2 A(\frac{\delta}{y})+B(\frac{\delta}{y})+C(\frac{\delta}{y})+D(\delta )-C(\delta )\}}}{\sqrt{f_{\gamma}(y,\delta )-f_{\gamma }(1,\delta )}}~-~1\right )dy\nonumber\\
&&~~~~~~~~~~~~~~~~~+1-\delta ^{-1}~+~\int _1^{\infty}
\left (1-e^{\frac{\gamma}{2}\{A(y^{-1})+B(y^{-1})\}}\right )dy\bigg \}.
\label{20}
\eea

The potential $V_1$ as a function of $L$ and $\Delta\Theta$ is obtained by inserting in (\ref{20}) the inversion of (\ref{17}),(\ref{17a}), i.e. $\delta =\delta (L,\Delta\Theta )$ and $l=l(L,\Delta\Theta )$. The qualitative features of this inversion problem are best understood by looking at the contour lines of $L,~\Delta\Theta $ and $V_1$ in the $(l^2,\delta ^4)$-plane. Since we are discussing string configurations with $U_T\leq U_0<\infty $ we have $0\leq \delta\leq 1$. In addition, the $\Delta\Theta $-formula requires $l^2\geq 0$, while the
$L$-formula needs $l^2\leq (1-\delta ^4)e^{\gamma \{A(\delta )+D(\delta )\}}$. Altogether
our set of formulae is applicable in the region
\footnote{Using the fact that $A(\rho )+D(\rho )$ as well as $C(\rho )-D(\rho )$ in $0\leq \rho\leq 1$ are negative and monotonously decreasing
one can convince oneselves that the integrals in (\ref{17}),(\ref{17a}) and (\ref{20}) are well defined $inside$ ${\cal G}$ for arbitrary $\gamma $. They exist even in a larger region, for the simplest case $\gamma =0$ in $0\leq \delta <1,~-\infty <l^2<2$.}
\beq
{\cal G}~=~\left \{(l^2,\delta ^4)~\vert ~~0\leq l^2\leq (1-\delta ^4)e^{\gamma \{A(\delta )+D(\delta )\}},~\delta ^4\geq 0 \right \}~.
\label{21}
\eeq   
A generic feature of these integrals is their monotonous increase with $l^2$ at
fixed $\delta $. As a consequence, $V_1$ and $\Delta\Theta $ are monotonously increasing functions of $l^2$. Due to the inverse behaviour of the square root prefactor it is difficult to make a definite statement concerning $L$.

The following discussion of the contour line maps is based on explicit analysis of the $\gamma =0$ case. However, the qualitative features, forced by the singularities of the functions under discussion, are not changed by switching on $\gamma $. We start with the $\Delta\Theta $ lines in $0\leq \delta ^4\leq1,~~0\leq l^2\leq 2$. Here $(l^2,\delta ^4)~=~(0,1)$ is a singular point \footnote{
Vanishing of the factor $l$ versus logarithmic divergence of the integral for $\delta ^4\rightarrow 1$.}. Across this point run all contour lines from $\Delta\Theta =0$ up to $\Delta\Theta =\infty $. The $\Delta\Theta =0$ line goes straight down to the point $(0,0)$, the $\Delta\Theta =\infty $ line first runs parallel to the $l^2$ axis to $(2,1)$ and then straight down to $(2,0)$. The $\Delta\Theta =\pi $ line goes through $(l^2,\delta ^4)=(1,0)$.

Let us turn to the $L$-lines. Again $(0,1)$ is a singular point. Here the
$L=\infty $ line at $\delta ^4=1,~~-\infty <l^2<0$ meets the straight zero-line
connecting $(1,0)$ with $(0,1)$. The approach of all lines with $0\leq L<\infty $
to the singular point is possible due to the balance between the logarithmic
divergent integral for $\delta ^4\rightarrow 1$ and the vanishing square root
prefactor. Therefore, $all$ lines with $L>0$ start tangent to the $L=\infty $
line into the region $\l^2<0$. Those with small enough $L$ must turn to the
$l^2>0$ region, since they must stay close to the zero line also away from the
singular point. Altogether $L$-lines up to some maximal value $L_{\mbox{\scriptsize max}}$ cross the $\delta ^4$-axis twice. For $l^2=0$ the inversion of (\ref{17}) exists only for $L\leq L_{\mbox{\scriptsize max}}$, for the $\gamma =0$ case see also \cite{theisen,brand}. From the point where the $L_{\mbox{\scriptsize max}}$-line touches the $\delta ^4$-axis to the point $(l^2,\delta ^4)=(1,0)$
extends a ridge of decreasing height. For a given $0\leq l^2<1$ the inverse of (\ref{17}) exists for values of $L$ smaller than the corresponding ridge height, 
only. 

Finally, the $V_1$ map is governed by a singularity with crossing
contour lines at $(l^2,\delta ^4)=(1,0)$. On the $l^2$-axis $V_1$ tends to
$\infty (-\infty )$ for $l^2>1(<1)$. Also $l^2=2,~~0\leq\delta ^4<1$ is a line of 
$V_1=\infty $. On the contrary, $V_1$ takes finite values on $\delta ^4 =1,~~l^2<2$. In particular one finds $V_1=0$ at $(0,1)$. The zero line of $V_1$ from
$(0,1)$ to $(1,0)$ is the critical line we are after. Beyond this critical line
there is no longer any force between the $Q\bar Q$-pair. From the analysis
of \cite{theisen,brand} and a theorem in \cite{mina} we know that, for $\gamma
=0$, the zero line coming from $(0,1)$  at first must go to the $l^2<0$ region
to cross the $\delta ^4$-axis at a value below the $L$-ridge. We have no general proof that the $V_1=0$ line is below the $L$-ridge in whole ${\cal G}$, but
there is numerical evidence for this scenario for not too large $\gamma $.\\

%%%%%%%%%%%%%%%%%%%%%%%%%%%%%%%%
\begin{figure}
\begin{center}
\mbox{\epsfig{file=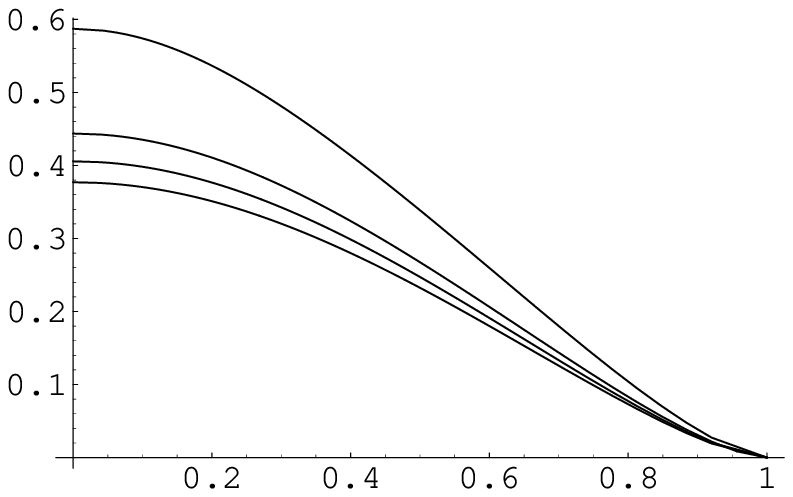 , width=90mm}}
\end{center}
\noindent {\bf Fig.1}\ \ {\it
Critical lines $V_1=0$ as functions of $\frac{1}{\pi}\cdot\Delta\Theta $
(horizontal) and $\frac{\pi}{2}T\cdot L$ (vertical). The lowest line is for
$\gamma =0$, the others for $\gamma = 0.01$, $0.02$, and $0.05$  respectively.}
\end{figure}
%%%%%%%%%%%%%%%%%%%%%%%%%%%%%%%%%%%%%%%%%%%%%%%%%%%%%%%%%%%%%%%
After this qualitative discussion of the behaviour of the potential we
present the results of a numerical evaluation. In fig.1 the critical lines,
$(V_1=0)$, in the $(\frac{1}{\pi}\cdot\Delta\Theta ~,\frac{\pi}{2}T\cdot L)$-plane, obtained for various values of $\gamma $ are shown. Beyond the critical line there is no $Q\bar Q$ force. Although for $\gamma =0$, i.e. the zeroth
order metric used in \cite{theisen,brand}, the potential is $R$-dependent,
the $V_1=0$ contour line in the $(\Delta\Theta ~,L)$-plane is independent of
$R$. With the first order corrected metric of \cite{tsey,paw} the $R$-dependence
originates from the $\gamma $-dependence via (\ref{4}). Obviously, the correction acts in the right direction, the critical line for the onset 
of total screening is shifted to larger distances over the whole range of $\Delta\Theta $. The chosen
values of $\gamma = 0.01$, $0.02$ and $0.05$ correspond to $R^2=2.468$, $1.959$ and $1.443$ respectively.
\newpage
\noindent
%%%%%%%%%%%%%%%%%%%%%%%%%%%%%%%%%%%%%%%%%%%%%%%%%%%%%%%%%%%%%%%
\begin{figure}
\begin{center}
\mbox{\epsfig{file=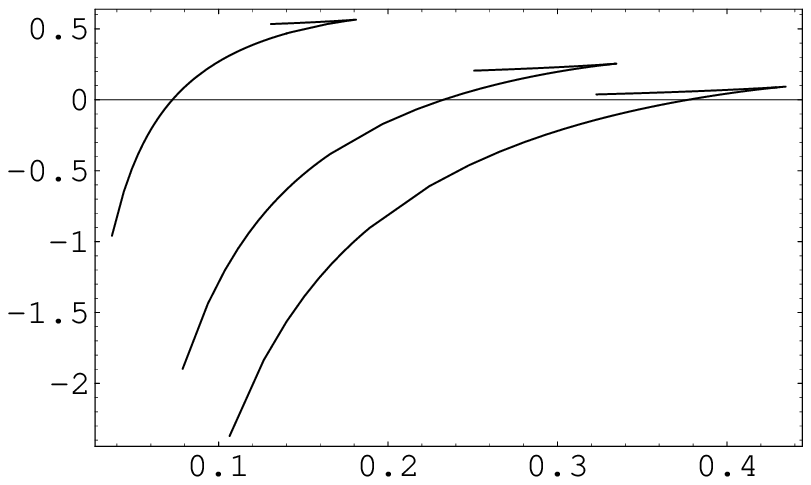, width=100mm}}
\end{center}
\noindent {\bf Fig.2}\ \ {\it
$\frac{1}{R^2T}\cdot V_1$ as a function of $\frac{\pi}{2}T\cdot L$ at $\gamma =0$ for various values of $\frac{1}{\pi}\Delta\Theta $, from left to right:
0.8, 0.5, 0.}
\end{figure}
%%%%%%%%%%%%%%%%%%%%%%%%%%%%%%%%%%%%%%%%%%%
To get an impression of the influence of $\Delta\Theta $ on the potential
we show in fig.2 $\frac{1}{R^2T}\cdot V_1$ as a function of $\frac{\pi}{2}T\cdot L$ for various values of $\frac{1}{\pi}\cdot \Delta\Theta $ in the case
of zeroth order metrics, i.e. $\gamma =0$. The $Q\bar Q$ force is zero for distances larger than the value of the zero of the drawn potential. For illustration we also followed the potential a little bit for positive values. The cusp
with the accompanying backward movement indicates that one goes beyond the
$L$-ridge discussed above.

For a better understanding of the mechanism switching off the
potential completely for $\Delta\Theta \rightarrow \pi$ we show (for $\gamma =0$) in fig.3 the $(l^2,~\delta ^4)$-plane with the region ${\cal G}$, the contour line
$V_1=0$ and contour lines of $\Delta\Theta $. Any $\Delta\Theta $-line below
$\pi $ enters ${\cal G}$ and the region of nonvanishing force for small enough
values of $\delta ^4$. The $\Delta \Theta =\pi $-line touches ${\cal G}$
at the point $(1,0)$ only. This figure also illustrates some statements made
in the qualitative discussion above.

Finally, fig.4 shows, again for $\gamma =0$, some contour lines of $\frac{\pi}{2}T\cdot L$ in the $(l^2,\delta ^4)$-plane. The numerical curves illustrate
the general analytic arguments given above. The $L$-ridge is visible clearly.
The $L$-line touching the $\delta ^4$ axis corresponds to $L_{\mbox{\scriptsize max}}=0.435$.\\
%%%%%%%%%%%%%%%%%%%%%%%%%%
\begin{figure}
\begin{center}
\mbox{\epsfig{file=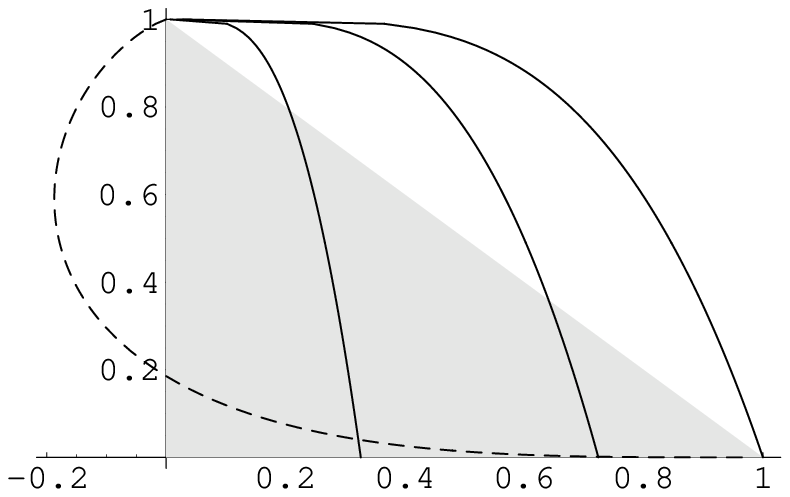, width=80mm}}
\end{center}
\noindent {\bf Fig.3}\ \ {\it $(l^2,\delta ^4)$ plane with region ${\cal G}$
for $\gamma =0$ (shadowed). Contour lines of $\frac{1}{\pi}\cdot \Delta\Theta
$ are shown for 0.5, 0.8, 1, from left to right.
The dashed line is the $V_1=0$ contour line.}
\end{figure}
%%%%%%%%%%%%%%%%%%%%%%%%%%%%%%%%%%%%%%%%%%%%%%%%%%%%%%%%%%%%%%%
%%%%%%%%%%%%%%%%%%%%%%%%%%
\begin{figure}
\begin{center}
\mbox{\epsfig{file=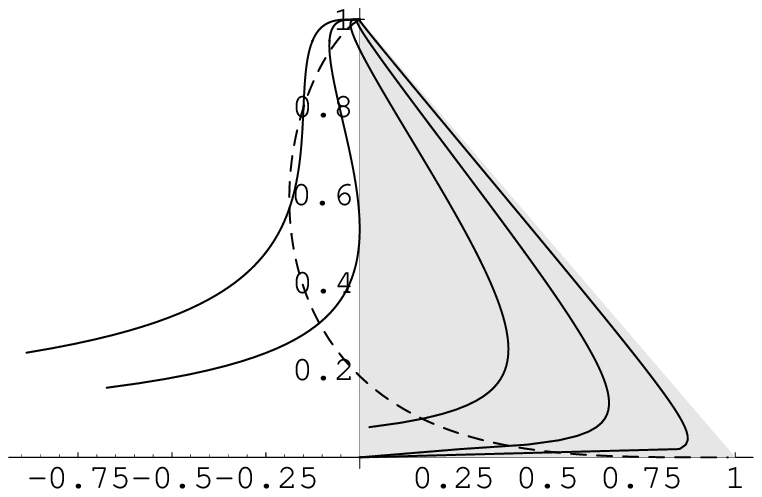, width=90mm}}
\end{center}
\noindent {\bf Fig.4}\ \ {\it $(l^2,\delta ^4)$ plane with region ${\cal G}$
for $\gamma =0$ (shadowed). Contour lines of $\frac{\pi}{2}T\cdot L$ are shown
for 0.5, 0.435, 0.3, 0.2, 0.1, from left to right. The dashed line is the $V_1=0$ contour line.}
\end{figure}
%%%%%%%%%%%%%%%%%%%%%%%%%%%%%%%%%%%%%%%%%%%%%%%%%%%%%%%
\newpage
\noindent
Closing this section we present the results of an expansion of
(\ref{17}),(\ref{17a}),(\ref{20}) for small $\delta $. It is related to
either $T\rightarrow 0$ at fixed distance or to small distances at fixed
temperature. Within this expansion one encounters the following $l^2$-dependent
integrals, expressible by the hypergeometric function $F$, the elliptic
integrals $K$ and $E$ as ($n\geq 0$)
\bea
b_{2n}(l)&=&\int ^{\infty}_{1}\frac{dy}{y^{2n}\sqrt{(y^2-1)(y^2+1-l^2)}}~=
~\frac{1}{2}B(n+\frac{1}{2},\frac{1}{2})~F(n+\frac{1}{2},\frac{1}{2},n+1,l^2-1)~,
\nonumber\\
b_{-2}(l)&=&\int _{1}^{\infty}\left (\frac{y^2}{\sqrt{(y^2-1)(y^2+1-l^2)}}~-~1
\right )dy\nonumber \\
&=&1+\frac{\pi}{2\sqrt{2-l^2}}F(\frac{1}{2},\frac{1}{2},1,\frac{1-l^2
}{2-l^2})-\frac{\pi}{2}\sqrt{2-l^2}F(-\frac{1}{2},\frac{1}{2},1,\frac{1-l^2}
{2-l^2})~.
\label{22}
\eea
All $b(l)$ are monotonously increasing with $l^2$.
Special values are
\bea
b_{2n}(0)&=&\frac{1}{4}B(\frac{n}{2}+\frac{1}{4},\frac{1}{2})~,\nonumber\\
b_{-2}(0)&=&1~+~\frac{1}{\sqrt{2}}\left ( K(\frac{\sqrt{2}}{2})-2E(\frac{\sqrt{2}}
{2})\right )~\approx ~ 0.401~,\nonumber\\
b_{2n}(1)&=&\frac{1}{2}B(n+\frac{1}{2},\frac{1}{2})~,\nonumber\\
b_{-2}(1)&=&1~.
\label{23}
\eea
Then we get $(c(\gamma )=\int _1^{\infty}(1-e^{\frac{\gamma}{2}(A(\frac{1}{x})+
B(\frac{1}{x}))})dx)$
\beq
\frac{V_1}{R^2T}~=~\frac{1}{1+15\gamma }\left (\frac{b_{-2}(l)-1}{\delta}
+1+c(\gamma )-\frac{1+75\gamma}{2}b_2(l)~\delta ^3~+~O(\delta ^7)\right )~,
\label{24}
\eeq
\bea
L\cdot T&=&\frac{2}{\pi}(1+15\gamma )\sqrt{1-l^2-(1+75\gamma )\delta ^4+O(\delta ^8)}\nonumber\\
&&~~~~~~~~~~~~~~~~\cdot\left (b_2(l)~\delta +\frac{1+75\gamma }{2}b_6(l)~\delta ^5+O(\delta ^9)\right )~,
\label{25}
\eea
\beq
\Delta\Theta ~=~2l\left (b_0(l)~+~\frac{1+75\gamma}{2}b_4(l)~\delta ^4~+~O(\delta ^8)\right )~.
\label{26}
\eeq
The insertion of inversed (\ref{25}),(\ref{26}) into (\ref{24}) is straightforward and will not be discussed here. 
%%%%%%%%%%%%%%%%%%%%%%%%%%%%%%%%%%%%%%%%%%%%%%%%%%%%%%%%%%%%%
\section{Spacelike Wilson loops}
To describe a string world sheet satisfying the boundary condition (\ref{7})
it is convenient to choose the gauge
\beq
\tau ~=~x^2,~~~~\sigma ~=~x^1~.
\label{27}
\eeq
The symmetry of the problem allows to restrict ourselves to world surfaces
with\\$\partial _{\tau}x^M=\delta _2^M$, $\partial _{\sigma}x^0=
\partial _{\sigma}x^3=0$. Then the action becomes, compare (\ref{9}),
\bea
S_{\vert \mbox{\scriptsize case 2}}&=&\frac{Y}{2\pi}\int _{-\frac{L}{2}}
^{+\frac{L}{2}}d\sigma ~e^{\frac{\gamma}{2}C(\frac{U_T}{U})}\nonumber\\
&\cdot &\sqrt{\frac{U^4}{R^4}~e^{\gamma C(\frac{U_T}{U})}+~U^{\prime \,2}\frac{U^4}{U^4-U_T^4}~e^{\gamma B(\frac{U_T}{U})}~+~\Theta ^{\prime \,2}U^2~e^{\gamma D(\frac{U_T}{U})}}~.
\label{28}
\eea
The conserved quantities $\tilde C_1$ and $\tilde C_2$ defined analogously
to (\ref{12}) will be parametrised by
\bea
\tilde C_2&=&U_0~l\nonumber\\
\tilde C_1&=&-\frac{U_0^2}{R^2}~e^{\gamma C(\delta )}~\sqrt{1-l^2e^{-\gamma \{C(\delta )+D(\delta )\}}}~.
\label{29}
\eea
Then we find with the notation
\beq
g_{\gamma}(y,\delta )~=~y^4~e^{2\gamma \{C(\frac{\delta}{y})-C(\delta )\}}
-y^2l^2~e^{\gamma \{ C(\frac{\delta}{y})-D(\frac{\delta}{y})-2C(\delta )\}}~,
\label{29a}
\eeq
repeating the steps in section 2,
\beq
L\cdot T~=~\frac{2}{\pi}~(1+15\gamma )~\sqrt{g_{\gamma}(1,\delta )}~\delta\int _1^{\infty}
\frac{e^{\frac{\gamma}{2}\{B(\frac{\delta}{y})-C(\frac{\delta}{y})\} }~dy}{\sqrt{y^4-\delta ^4}\sqrt{g_{\gamma}(y,\delta )-g_{\gamma}(1,\delta )}}~,
\label{30}
\eeq
\beq
\Delta\Theta~=~2l~\int _1^{\infty} \frac{y^2~e^{\frac{\gamma}{2}\{B(\frac{\delta}{y})+C(\frac{\delta}{y})-2D(\frac{\delta}{y})-2C(\delta )\}}~dy}
{\sqrt{y^4-\delta ^4}\sqrt{g_{\gamma}(y,\delta )-g_{\gamma}(1,\delta )  }}~,
\label{31}
\eeq
\beq
\frac{V_2}{R^2T}~=~(1+15\gamma )^{-1}\bigg \{\delta ^{-1}~\int _1^{\infty}
\left (\frac{y^4~e^{\frac{\gamma}{2}\{B(\frac{\delta}{y})+3C(\frac{\delta}{y})
-2C(\delta )\}}}{\sqrt{y^4-\delta ^4}\sqrt{g_{\gamma}(y,\delta )-g_{\gamma}(1,\delta )} }-1\right )~+~1~-~\delta ^{-1}\bigg \}~.
\label{32}
\eeq

To define the renormalised potential $V_2$ we have chosen as the subtraction
$\frac{\Lambda -U_T}{\pi}$. Contrary to the previous section the pssc configuration has a $L$-dependent action. This is due to different behaviour on
the horizon, $G_{22}\neq 0$ versus $G_{00}=0$. With the same subtraction
as for $V_2$ one gets, e.g. for $\Delta\Theta =0$,
\beq
V_{2,\mbox{\scriptsize pssc}}~=~\frac{U_T^2}{2\pi R^2}~e^{\gamma C(1)}\cdot L~+~
\frac{U_T}{\pi}\int _1^{\infty}\left (\frac{y^2e^{\frac{\gamma}{2}\{B(\frac{1}{y})+C(\frac{1}{y})\}}}{\sqrt{y^4-1}}~-1\right )dy~.
\label{33}
\eeq
We expect this to be energetically disfavoured over the whole range of $L$.

The $L-\delta $ relation differs qualitatively from that in section 2.
Due to the absence of $\delta $ in the square root prefactor we can reach arbitrary large
values of $L$ by approaching the singularity of the integral at $\delta =1$.
Furthermore, $L$ is monotonously increasing with $\delta ^4$.
\footnote{This is obviously for $\gamma =0$. It can be demonstrated by
simple estimates also for small enough values $\gamma >0$.}
The leading terms in the expansion of all the integrals in (\ref{30})-(\ref{32}) for $\delta \rightarrow 1$ coincide due to the equality of the leading term of the expansion of the respective integrands for $y\rightarrow 1$.
Denoting with a fixed $\epsilon >0$
$$I(\epsilon ,l)~=~\int _1^{1+\epsilon}\frac{dy}{\sqrt{y^4-\delta ^4}\sqrt{
g_{\gamma}(y,\delta )-g_{\gamma}(1,\delta )}}~, $$
we get
\bea
L\cdot T&=&\frac{2}{\pi}~(1+15\gamma )~\sqrt{g_{\gamma}(1,1)}~e^{\frac{\gamma}{2}\{B(1)-C(1)\}}~(I(\epsilon ,l)~+~O(1))~,
\nonumber\\
\Delta\Theta&=&2l~e^{\frac{\gamma}{2}\{B(1)-C(1)-2D(1)\}}~ (I(\epsilon ,l)~+~O(1))~,\nonumber\\
\frac{V_2}{R^2T}&=&(1+15\gamma )^{-1}~e^{\frac{\gamma}{2}\{B(1)+C(1)\}}~(I(\epsilon ,l)~+~O(1))~.
\label{34}
\eea
Using the first line of (\ref{34}) to eliminate $I$ in the last line one
arrives at
\beq
V_2~=~\frac{\pi R^2T^2}{2}(1-\frac{265}{8}\gamma )\cdot \frac{L}{\sqrt{1-l^2e^
{\frac{5}{4}\gamma}}}~+~O(1)~.
\label{35}
\eeq
At the very end we want to know the potential for fixed $\Delta\Theta$.
According to the second line of (\ref{34}) at $L\rightarrow\infty$ this
requires $l\rightarrow 0$. Using the first line of (\ref{34}) to eliminate $I$
in the second line one gets $l=\frac{\Delta\Theta }{\pi LT}e^{\gamma (15+D(1))}
+o(\frac{1}{L})$ and altogether up to linear terms in $\gamma $
\beq
V_2(\Delta\Theta ,L)~=~\frac{\pi R^2T^2}{2}(1-\frac{265}{8}\gamma )\cdot L~+~
\frac{1}{4\pi}R^2(\Delta\Theta )^2(1+\frac{15}{8}\gamma )\cdot \frac{1}{L}~+~O(1)~.
\label{36}
\eeq 

The terms which we neglected and indicated as $O(1)$ are, up to eq.(\ref{35}),
more precisely $\propto Le^{-\pi TL}$,  as follows by repeating the estimates of ref.\cite{olesen}. There the case $\Delta\Theta =0$ was studied and the absence of any L\"uscher term $\propto 1/L$ in the large $L$ expansion
was pointed out to be an obstacle for the potential, derived in the 
approach of \cite{malda,rey,witten}, to be identified with a QCD potential.
We find it interesting that for $\Delta\Theta\neq 0$ an $1/L$ term
arises. However, due to the wrong sign and a coupling dependence
\footnote{We thank P. Olesen for a comment on this fact.} it cannot be
interpreted as a L\"uscher term. One the other side one should keep
in mind that this term is also not Coulombic since in 2+1 dimensions the Coulomb term is $\propto \log L$.
%%%%%%%%%%%%%%%%%%%%%%%%%%%%%%%%%%%%%%%%%%
\section{Conclusions}
We showed that within the classical approximation the static quark-antiquark
potential is completely switched off for antipodal $Q\bar Q$ orientation
in the internal $S^5$ also for $T>0$. In contrast to the $T=0$ case this
is, due to the broken SUSY, $not$ required by any BPS argument. The effect
is present both for the zeroth order background metrics of refs.\cite{theisen,brand} as well as for the higher curvature corrected background of ref.\cite{tsey,paw}. One should add, that certainly this is an effect of the backgrounds used.
In other circumstances, with BPS arguments set out of work, nonvanishing
$Q\bar Q$ forces may be possible for all orientations in internal space. Here,
for example, we have in mind  a ${\cal N}=1,~~T=0$ case based on a simple orbifold construction over $S^5$ \cite{kachru}.

The replacement of the lowest order background by the corrected ones shifts the
critical line in the orientation-distance plane to larger distances for all
values of the relative orientation in internal space. This shift acts in the right direction, if one follows the arguments of ref.\cite{gross} stating
the vanishing of the total screening as soon as all corrections are taken into account. 

From the study of spacelike Wilson loops we got the string tension for a temperature zero gauge theory in (2+1) dimensions. This tension is independent
of the relative internal orientation of the quarks. However, the tension is
sensible to switching on the string target space background corrections.
The subleading terms in a large distance expansion of the potential
are dependent on the relative internal orientation. In particular
we found a nextleading $1/L$ term, absent for parallel orientation.

In this picture $T$ plays the role of a regularization parameter which, to make
contact with (2+1)-dimensional non-supersymmetric QCD, has to be sent to infinity. To keep a finite tension, $R^2$ has to go to zero. Obviously, our approximation is not
suited to study this limit. However, our formula (\ref{36}) suggests that
the $\Delta\Theta $ dependent $1/L$ term drops out in this limit. This
indeed should happen since there is no place in QCD for the internal $S^5$ orientation parameter which is a remnant of ${\cal N}=4$ SUSY.\\[10mm]
%%%%%%%%%%%%%%%%%%%%%%%%%%
{\bf Acknowledgement:} 
We would like to thank A. Karch, V. Pershin, Chr. Preitschopf and S. Theisen
for useful discussions.
%%%%%%%%%%%%%%%%%%%%%%

\end{document}